\documentclass[preprint,showpacs,preprintnumbers,apl]{revtex4}
\usepackage{graphicx}

\begin{document}

\newcommand {\be}{\begin{equation}}
\newcommand {\ee}{\end{equation}}
\newcommand {\bea}{\begin{eqnarray}}
\newcommand {\eea}{\end{eqnarray}}
\newcommand {\nn}{\nonumber}

\title{Electron-phonon bound states and impurity band formation 
in quantum wells}
\author{Bruna P. W. de Oliveira}
\author{Stephan Haas}
\affiliation{Department of Physics and
Astronomy, University of Southern California, Los
Angeles, CA 90089}

\date{\today}

\begin{abstract}
A generalized propagation matrix method is used to study how  
scattering off local Einstein phonons affects resonant
electron transmission through quantum wells. In particular, the parity and 
the number of the phonon
mediated satellite resonances are found to depend on the available 
scattering channels. For a large number of phonon channels,
the formation of low-energy impurity bands is observed. 
Furthermore, an effective theory is developed which 
accurately describes the phonon generated 
sidebands for sufficiently small electron-phonon coupling.
Finally, the current-voltage characteristics caused by phonon assisted 
transmission satellites are discussed for a specific double barrier geometry.  
\end{abstract}

\maketitle

\section{Introduction}

Resonant tunneling through quantum wells has been 
extensively studied in semiconductor heterostructures,
such as $\rm GaAs/Al_xGa_{1-x}As$ double barriers
\cite{goldman87,boebinger90,chen91,geim94,kim03}. More recently, 
analogous electron transmission processes have
also been investigated in the context of 
molecular junctions \cite{nitzan03,benesch06,mitra04,braig03,galperin07} and 
mesoscopic rings \cite{yacoby95,buks96}.
Resonant tunneling is a purely quantum effect whereby 
electrons pass through structures made of
potential wells and barriers with unit or 
near-unit transmission probabilities if they enter the 
quantum well at the particular energies of the structure's bound states.  
Following the initial experimental observation of satellite peaks
of these transmission 
resonances \cite{goldman87}, a large volume of theoretical work 
\cite{levi89,cai89,cai90,cai91,wu91,stovneng91,bagwell92,grein93,zhou93,bonca95,bonca97,haule99,mohaidat93,brandes02} has focused on the effects of
phonon scattering on the electronic tunneling. Early on, it was 
recognized that perturbative treatments tend to miss the essential 
feedback effects between elastic and inelastic channels which lead 
to these satellite features in the electronic transmission 
 \cite{levi89}. In particular, it was found 
that electron-phonon scattering processes can cause the formation
of polaronic bound states, leading to phonon assisted resonant 
tunneling
\cite{cai89,cai90,cai91,wu91}. More recent works have shown that within a 
tight-binding description 
these features are further enhanced \cite{stovneng91}, and phonon 
bands can form \cite{bonca95,bonca97}. Furthermore, theoretical models 
have been generalized to include the effects of the three-dimensional 
environment \cite{bagwell92,zhou93},
non-equilibrium dynamics \cite{grein93},
and finite temperatures \cite{haule99}.

In this paper, we examine the hierarchy of polaronic resonances in
the electron transmission through quantum well structures. In 
particular, we focus on even-odd effects with respect to the number 
of available phonon channels and on the emergence of impurity 
bands as this number becomes large. We also apply an effective 
theory which reproduces the dependence of the resonance peaks on 
the electron-phonon coupling strength and the phonon energy in the 
limit of sufficiently small coupling. The method we are using is 
a generalization of the propagation matrix technique \cite{levibook}
which takes into account elastic electron scattering at potential steps
as well as scattering off local Einstein phonons. This approach allows 
a numerically exact calculation of the electron transmission through
quasi-one-dimensional heterostructures without any perturbative 
constraints, such as limitations to particular parameter regimes,
or restrictions to specific energy ranges, such as low-energy 
resonant states. In addition, our method accounts for the 
feedback between an adjustable number of phonons and the elastic 
transmission channel, and is therefore suitable to accurately describe    
the interplay between non-perturbative resonances of the many-body 
system.

Before proceeding to the discussion of resonant tunneling through
specific semiconductor double wells, let us briefly point out 
some similarities and differences of this system 
with electronic transport through 
molecular junctions.\cite{nitzan03,benesch06,mitra04,braig03,galperin07}
In the theory of both physical systems, 
many-body methods are combined with scattering theory to obtain the
tunneling density of states and the resulting current-voltage
characteristics for electronic transport
through small objects with quantized energy levels. In both cases one
observes the formation of phonon assisted satellite features as
the electrons scatter off local vibrational modes.
However, there are 
several significant differences between these systems, as we will see below.
Electron transport through
layered semiconductor structures
exhibit resonant tunneling features which
coexist with continuum contributions. These are typically absent in
molecular transistors. Furthermore, since semiconductor 
hereostructures 
are manmade, the specific resonance levels can be controlled by
layer thickness and composition and are thus tunable. Moreover,
the experimental current-voltage curves for semiconductor 
heterostructures are quite
different from molecules, i.e. they show peak features rather than the
steps characteristic for molecular systems. 
\cite{semi}

\section{Model and Method}

We wish to determine the transmission probability of
electrons through potential structures of arbitrary profile, with the 
possibility of 
exciting local Einstein phonon channels. The 
basic Hamiltonian for this problem,
\bea
H=
\sum_k \epsilon (k) c^\dagger_k c_k 
+ \sum_x V(x) c^\dagger_x c_x
+ \sum_{x_i} \hbar\omega b^\dagger_{x_i} b_{x_i}
+ g\sum_{x_i,k,k'} \delta(x-x_i)
\left(b^\dagger_{x_i}+b_{x_i} \right)c^\dagger_k c_{k'} ,
\eea 
describes electrons with  creation and annihilation operators
denoted by $c^\dagger$ and $c$, and a dispersion $\epsilon(k) = 
\hbar^2 k^2/2m$, propagating through a potential structure whose
real-space profile is given by 
$V(x)$. In addition, local Einstein phonon scatterers with creation and
annihilation operators
$b^\dagger$ and $b$ and energy $\hbar\omega$ are placed at impurity 
sites $x_i$.  The electron-phonon interaction is controlled by the 
coupling constant $g$, which has units of energy times length. 
The particular systems we have in mind are layered semiconductor
heterostructures, such as $\rm GaAs/Al_xGa_{1-x}As$.
For these systems, the use of a momentum independent electron-phonon
coupling constant is standard, and can be viewed as a reliable lowest
order approach.\cite{brandes02} 

To find the electronic transmission probability we use the propagation 
matrix method, which is applied in the following way: for each step at 
position $j$ in the potential profile, we construct a propagation matrix 
$\hat\rho_{step}^j$, and in between neighboring steps at a distance $L_j$ 
apart, we construct a propagation matrix $\hat\rho_{free}^{L_j}$. The 
elements of $\hat\rho_{step}^j$ depend on the boundary conditions of the 
electronic wavefunction at the potential step at position $j$. The matrix 
$\hat\rho_{free}^{L_j}$ is diagonal, and its elements depend on the phase 
picked up by the electron as it propagates through a 
length $L_j$ between potential steps. The total propagation matrix is 
given by the product of the individual matrices:

\begin{equation}
\hat\rho=\hat\rho_{step}^1\hat\rho_{free}^{L_1}...\hat\rho_{free}^{L_N}\hat\rho_{step}^N.
\end{equation}

An example of the propagation matrix method is given in the appendix. For 
a system without phonon excitations, the propagation matrix is simply a 2 
$\times$ 2 matrix. When the electron excites phonons as it penetrates the 
structure, the propagation matrix grows as $(2n+2) \times (2n+2)$, where 
$n$ is the number of phonon channels. When several phonons are excited it 
becomes necessary to find the transmission 
probability of an electron as a function of energy numerically. The idea 
is to solve 
a system of linear equations of the form $\hat\rho\mathbf{x}=\mathbf{a}$, 
where $\mathbf{x}$ is the vector whose terms correspond to the 
transmission and reflection coefficients 
$\mathbf{x}=(t_0,r_0,...,t_n,r_n)$ and 
$\mathbf{a}=(a_0,b_0,...,a_n,b_n)$, where the coefficients $a_l$ and 
$b_l$ depend on the initial conditions of the problem. In our problem 
there is no reflection as the electron exits the potential profile, so we 
can set the reflection coefficients $r_l=0$ for all $l$, therefore 
reducing the number of equations in the system by half:

\begin{equation}
 \left(
\begin{array}{cccc} 
\rho_{11} & \rho_{13} & ... & 
\rho_{1n-1} \\ 
\rho_{31} & \rho_{33} & ... & 
\rho_{3n-1} \\
\vdots &\vdots &\ddots & \vdots
 \\ 
\rho_{n-11} & \rho_{n-12} & 
... & \rho_{n-1n-1} 
\end{array} \right) 
\left(
\begin{array}{c}
t_0\\
t_1\\
\vdots \\
t_n
\end{array} \right)
= \left(
\begin{array}{c}
a_0\\
a_1\\
\vdots \\
a_n
\end{array} \right).
\end{equation}
\\
Initially all phonons are in the ground state, and therefore 
$a_j=\delta_{j0}$. All that is left is to determine the $t_j$, and we can 
do so by solving the system above using a Gauss-Jordan elimination. Once 
these terms are found, we can calculate the transmission probability:


\begin{equation}
T(E)=\sum_{l=0}^n\frac{k_l(E)}{k_0(E)}|t_l(E)|^2 ,
\end{equation}
where $k_l$ is the momentum of the electronic wave function in a 
channel with $l$ phonons.
With this approach it is possible to plot transmission probability versus 
energy.

\section{Numerical Results}

\begin{figure}[h]
\centering
\includegraphics[scale=0.6]{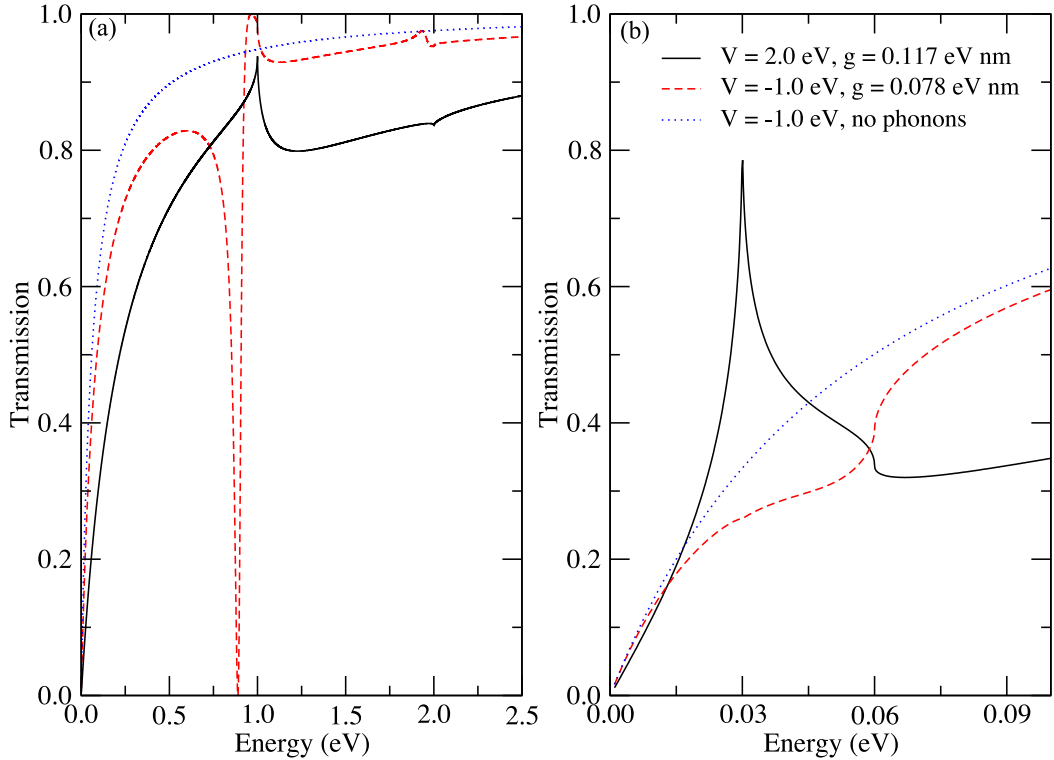}
\vspace{-1.6cm}
\caption{\small (a) Electron transmission probability 
through a delta potential. The solid black 
line represents a
repulsive potential with an electron-phonon 
coupling constant $g=0.117$ eV nm. The dashed red line corresponds to 
transmission through an attractive delta potential with
$g=0.078$ eV nm. In both cases, the electron is allowed to 
excite two local oscillator levels with energies
$\hbar\omega$ = 1.0 eV and $2.0$ eV. The blue line is for $g=0$.  
(b) Electron transmission probability through a finite-width
potential barrier/well of width $L=0.1$ nm.
The solid black
line represents a
repulsive rectangular potential of strength $V_0=2.0$ eV with 
$g=0.117$ eV nm. The dashed red line corresponds to
transmission through an attractive rectangular potential $V_0=-1.0$ eV with
$g=0.078$ eV nm. The local oscillator levels are chosen to be 
at $\hbar\omega$= 0.03 eV and 0.06 eV, positioned
at the centers of the rectangular potentials.
}
\end{figure}

As a test of the validity of the propagation matrix method, we first 
examine two cases which have previously been studied in the literature. 
In Fig. 1(a), we show the transmission through 
repulsive (black) and attractive (red) delta potentials, which allow 
the excitation of two local vibrational modes at 
$\hbar\omega$ = 1.0 eV and 2.0 eV. For comparison, we also show 
the case (blue) without coupling to these local Einstein modes. In their 
absence, the transmission increases monotonically with the energy of the 
incoming electrons. However, in the presence of inelastic scattering channels
resonance features in the form of spikes and dips in $T(E)$ 
occur at energies slightly below the local oscillator levels.
They indicate the formation of 
bound states\cite{cai90,bagwell92,brandes02},
manifested by Fano features which for 
attractive potentials can completely suppress 
electron transmission right below the resonance energy (red line in
Fig. 1(a)).
These features arise from the strong
feedback between inelastic and elastic scattering processes, and are
easily missed in perturbative treatments\cite{levi89}.
The parameters in Fig. 1 
have been chosen identical to previously published data
\cite{bagwell92,brandes02} to demonstrate full agreement of methods.

As shown in Fig. 1(b), the phenomenon of polaron-type bound state formation
persists for finite-width wells and barriers. 
\cite{cai90} Here, the location of the 
Einstein scatterers are chosen at the center of the rectangular
potential profiles. 
Bearing in mind experimentally relevant scales, we consider vibrational
energies two orders of magnitude lower than in the Fig. 1(a), i.e. at 
$\hbar\omega = 0.05$ eV and $0.10$ eV. In analogy to the case of
delta potentials, resonances are observed
at both energy levels. However, electron transmission is
suppressed with respect to the case of delta potentials because of
the finite spatial extent of the wells and barriers. 
 
\begin{figure}[h]
\centering
\includegraphics[scale=0.6]{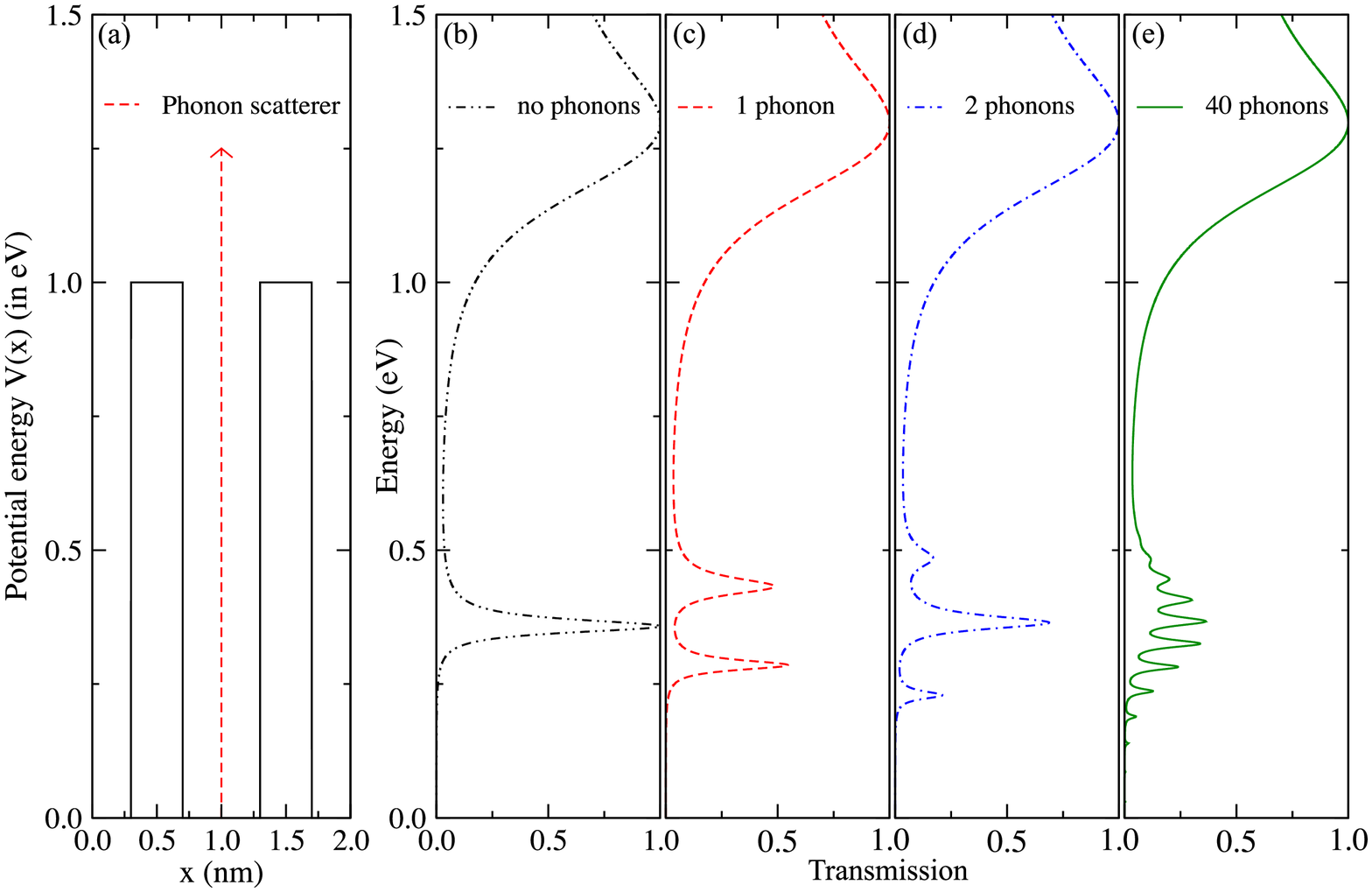}
\caption{\small (a) Symmetric double rectangular potential
with barrier width 0.4 nm, well width 0.6 nm, and barrier energy 
$V_0 =$ 1.0 eV. A local phonon scatterer is located at the 
center of the double barrier. (b-e) Electron transmission probabilities 
for various numbers of accessible phonon channels. 
(b) corresponds to the case without 
electron-phonon interaction ($g=0$), (c) represents 
the transmission coefficient for one phonon channel, 
(d) for 2 phonon 
channels, and (e) for 40 phonon channels. In 
(c-e) the electron-phonon coupling is set to $g=0.04$ eV nm, and the phonon 
frequency is $\hbar\omega=0.01$ eV. With increasing number of phonon 
channels one observes the formation of a band.}
\end{figure}

Next, we turn to the case of electron transmission through more
complex quantum well
structures. 
Focusing on symmetric potential profiles, let us consider rectangular 
double barriers of length $0.4$ nm, separation $0.6$ nm, and height 
$V_0$ = 1 eV. A local Einstein scatterer is placed at the 
center of the well, as illustrated in Fig. 2(a). The vibrational energies
are $\hbar\omega_n = n \hbar\omega$ with $\hbar\omega = 0.01$ eV and $n = 1, 2, 3, ...$.
In the absence of phonon scattering, shown in Fig. 2(b), one observes  
a bound state at $E=0.358$ eV which allows resonant tunneling 
with unit transmission. In the following, we examine the effects 
of inelastic scattering on this resonant feature.  
In the presence of a phonon scatterer with a single available inelastic 
channel at $\hbar\omega_1 = 0.01$ eV (Fig. 2(c)), the bound state is split
into two satellites, separated by approximately equal energy gaps with 
respect to the energy of the original bound state. Such ``side bands" 
have been the focus of numerous earlier studies. 
\cite{kim03,cai89,cai90,cai91,wu91,stovneng91,bagwell92,zhou93,bonca95} 
Note that for the semiconductor double barrier structures studied
here the magnitudes of the energy splits between these phonon 
assisted satellite peaks are considerably larger than the weak coupling result,
$E_0 \pm \hbar\omega_1$, where $E_0$ is the energy of the 
resonance in the absence of inelastic scattering. This is due to strong 
renormalization of the bare electron-phonon coupling constant by the 
confinement of the electron wave function to the small well region, which 
will be discussed in more detail later on.

Here, we 
wish to examine how such phonon assisted satellite features merge into an
impurity band with increasing number of available inelastic channels. 
The generalized propagation matrix method is particularly suited for this 
task, as the propagation matrix for the system only increases linearly 
with the number of added vibrational modes. 
The pattern which emerges from Fig. 2 is that the
bound state splits into $n+1$ peaks, where $n$ is the number of phonon
channels which are excited. For instance, in the case of one excited 
phonon with energy $0.01$ eV and electron-phonon coupling $\rm g=0.04 eV\cdot nm$, we 
find the peaks to be at positions $E_1=0.2833$ eV and $E_2=0.4340$ eV
(Fig. 2(c)), 
which differ from the zero-phonon case (Fig. 2(b))
by $\Delta_1=-0.0747$ eV and 
$\Delta_2=0.076$ eV. For two phonons (Fig. 2(d)), one finds    
3 peaks at $E_1=0.2255$ eV, $E_2=0.3645$ eV, and $E_3=0.4875$ 
eV, giving shifts of $\Delta_1=-0.1325$ eV, $\Delta_2= 0.0065$ eV and $\Delta_3=0.1295$ eV. 
This observation points to an interesting even-odd 
effect, whereby for odd numbers of phonon channels, there exists a  
central, non-bonding, peak, whereas for even numbers of phonon 
channels it is absent. It also implies that the satellites at
$E_n \approx
E_0 \pm \hbar\omega_n$ are bonding/anti-bonding pairs. Before investigating 
this aspect of multi-phonon-assisted resonant tunneling more closely, 
let us point out that in
the limit of many phonon channels (Fig. 2(e)) a low-energy band 
emerges. Note that the asymmetry in this impurity band is already 
anticipated in the asymmetry of the satellites for the few-phonon cases  
\cite{stovneng91,bagwell92}.

\begin{figure}[h]
\centering
\includegraphics[scale=0.6]{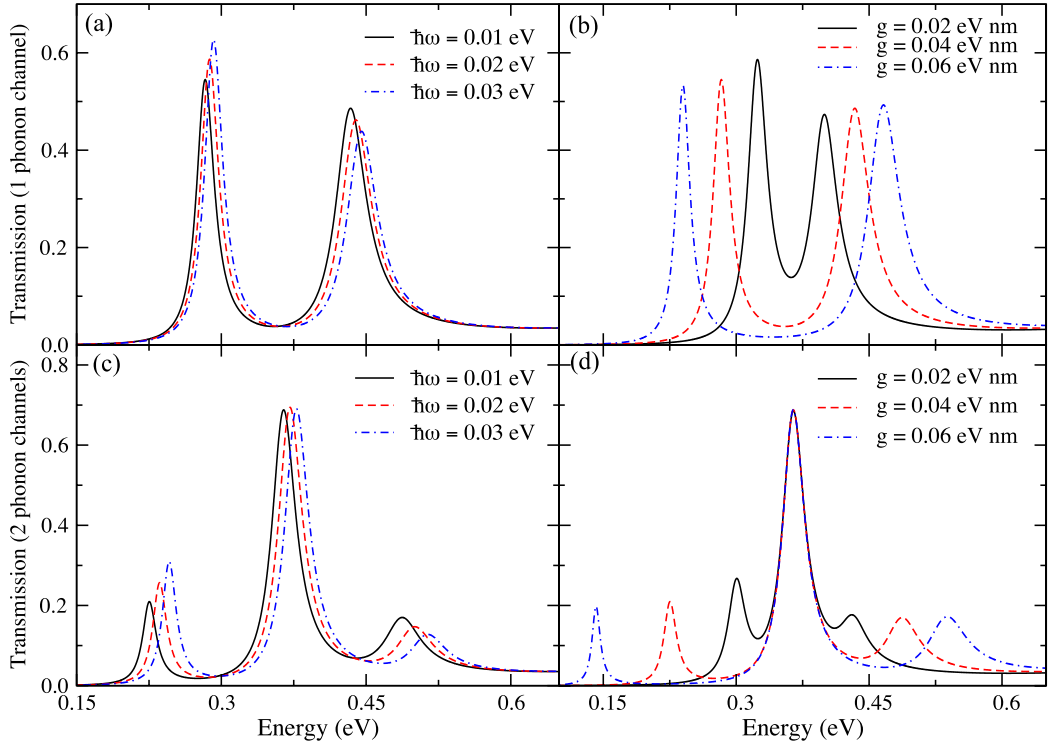}
\vspace{-1cm}
\caption{\small Electron transmission resonances for a 
double rectangular potential barrier with a phonon scatterer located at the
center. The parameters are chosen identical to Fig. 2, unless otherwise
specified. 
(a) and (b) correspond to a system with one phonon channel, and (c), (d) to a 
system with two phonon channels. In (a) and (c) the electron-phonon
scattering strength is kept constant at $g$ = 0.04 eV nm, and the phonon energy
$\hbar\omega$ is varied. One observes that 
the entire spectrum shifts to larger values of energy as 
$\hbar\omega$ increases. In (b) and (d) $\hbar\omega$ = 0.01 eV and $g$ is varied. 
In this case the
gaps between the transmission resonance peaks widen with
with increasing $g$. For two phonon channels, the  
central peak does not shift as $g$ is varied. }
\end{figure}

In Fig. 3 we examine the effects of the electron-phonon 
coupling and the vibrational energies on resonant transmission
through the same double barrier potential shown 
in Fig. 2(a), i.e. we focus on the low-energy transmission peaks.
The case of one available phonon channel is studied in 
Figs. 3(a) and (b), and the case of
two phonon channels is illustrated in Figs. 3(c) and (d).
Let us first keep the electron-phonon coupling constant at 
$\rm g=0.04 eV\cdot nm$,
and vary the energy of the vibrational levels. As observed in 
Figs. 3(a) and (c), increasing 
values of $\hbar\omega$ cause the entire spectrum of transmission 
resonances to shift to higher 
energies, whereas the gaps between the peaks remain constant. 
If instead we keep the vibrational energies fixed, i.e. 
at $\hbar\omega=0.01$ eV,
and vary $g$, the gaps between 
peaks are found to increase as $g$ increases (see Table 1).
Note that for the case of even numbers of
phonon channels the central non-bonding peak does not shift
with increasing $g$, whereas as the 
bonding/antibonding peaks move to higher and lower energies respectively.

\begin{table}[ht] 
\caption{Gap between bonding and antibonding peaks for different values of 
electron-phonon coupling $g$} 
\centering      
\begin{tabular}{|c| c| c| c|}  
\hline                        
$\rm g (eV \cdot nm)$ & 0.02 & 0.04 & 0.06  \\   
\hline
$\Delta$ (1 phonon) & 0.074 eV & 0.147 eV & 0.221 eV  \\ 
$\Delta$ (2 phonons) & 0.104 eV & 0.209 eV & 0.314 eV  \\ [1ex]       
\hline     
\end{tabular} 
\label{table:nonlin}  
\end{table} 

The observation of bound state energy splitting when there are Einstein
phonon channels in the system is analogous to degeneracy breaking in 
the linear Stark effect. For sufficiently 
small electron-phonon coupling, we can compute 
the first order energy shift quantitatively by treating the phonon energy 
and electron-phonon interaction terms in the Hamiltonian (1) as  
perturbations and by using time-independent degenerate perturbation theory 
to calculate the resulting energy shifts. The unperturbed eigenstates are 
denoted by
$|x, n\rangle$, where $n$ is the phonon quantum number and $x$ denotes the 
electron position in the well region of the potential profile $x\in[0,L]$.
In order to make analytical progress, the electron wave function is 
approximated by the infinite well wavefunction $\Psi(x)\approx \text{sin}(\pi 
x/L)/\sqrt{L}$, where $L$ is the length of the well. The 
resulting perturbation matrix has elements 
$\langle x,n|\sum_{x_i} \hbar\omega b^\dagger_{x_i} b_{x_i}
+ g\sum_{x_i,k,k'} \delta(x-x_i)
\left(b^\dagger_{x_i}+b_{x_i} \right)c^\dagger_k c_{k'} |x,n\rangle$,
which for  
the case of one phonon channel yields the $2\times 2$ 
perturbation matrix 
\begin{equation}
\hat{P}= \left(
\begin{array}{cc} 
0 & g/L \\ 
g/L & \hbar\omega \\ 
\end{array} \right).
\end{equation}
Assuming that the impurity site is located at $x_0=L/2$,
we have $\text{sin}(\pi 
x_0/L)=1$. The energy shifts are calculated by diagonalization of the 
perturbation matrix and are given by 
\bea
\lambda=\frac{\hbar\omega}{2}\pm\sqrt{\left(\frac{\hbar\omega}{2}\right)^2
+\left(\frac{g}{L}\right)^2}.
\eea
In practice, the well width $L$ can be made rather small, even compared 
to the scale of molecular junctions. This can lead to a significant
renormalization
of the electron-phonon coupling constant, $g \rightarrow g/L$, which in turn 
explains the the relatively large energy gaps between the phonon assisted 
satellites, observed in the propagation matrix results.  

To further illustrate how the bound state energies depend on the phonon 
energy and the electron-phonon coupling, we plot the lower 
and higher bound state energies as a function of $\hbar\omega$ (Figs. 
4(a) and 4(c)), and as a function of $g$ (Figs. 4(b) and 4(d)). 

\begin{figure}[h]
\centering
\includegraphics[scale=0.57]{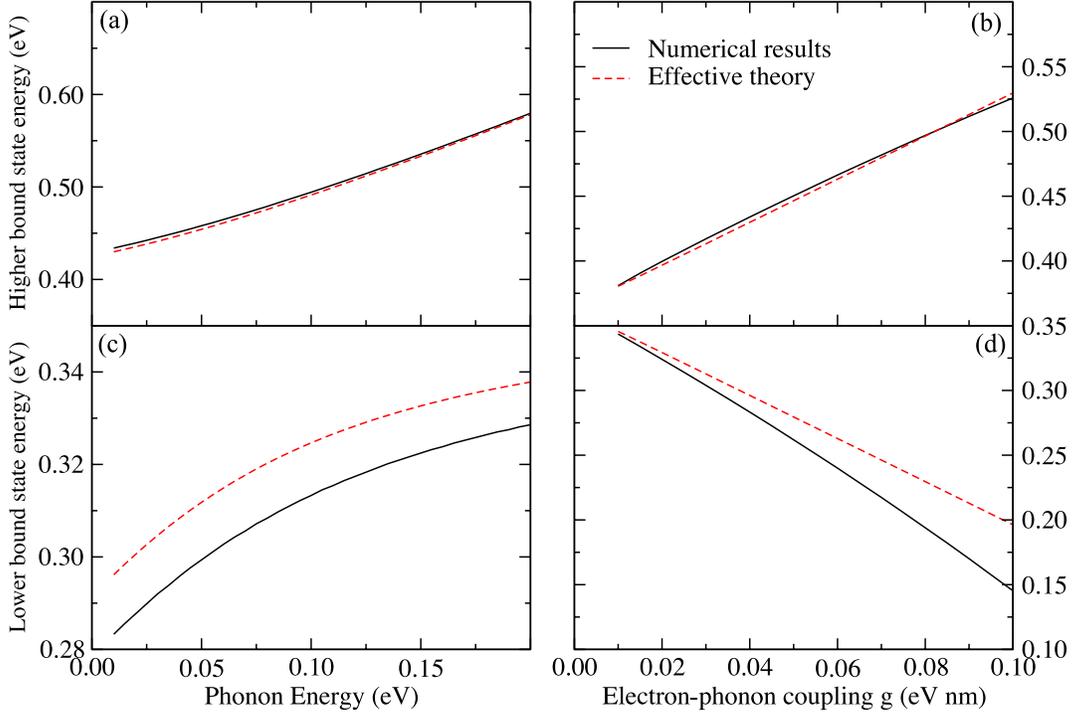}
\vspace{-1cm}
\caption{\small Bound state energies
in a double rectangular potential barrier with one phonon 
scatterer at the center (as depicted in Fig. 2(a)). For the case of one phonon channel, the 
energies of the two resulting transmission resonance peaks
are plotted as 
functions of the electron-phonon coupling constant $g$ 
and of the phonon energy $\hbar\omega$.
Exact numerical results (solid black lines) are compared with the effective
theory (dashed red lines) described in the text. 
In (a) and (c) the electron-phonon coupling is kept constant at
$g=0.04$ eV nm, and in 
 (b) and (d) the phonon energy is kept constant 
at $\hbar\omega=0.01$ eV.
In (a) and (b), for the higher bound state energies, the effective theory reproduces remarkably well the numerical results. 
In (c), the effective theory and numerical results are off-set by 
approximately $0.02$ eV. In (d), the effective theory matches the
numerical results for small values of $g$. For instance, when $g=0.04$ eV nm, the effective theory overestimates the lower bound state energy by $0.01$ eV.  }
\end{figure} 

The accuracy of the effective theory compared to the numerical results of 
the full propagation matrix calculation is striking, in particular for 
predicting the 
higher bound state energy (Figs. 4(a) and 4(b)). For the lower bound 
state, the accuracy increases for smaller $g$ (Fig. 4(d)), although for a 
fixed value of $\rm g=0.04 eV\cdot nm$ and variable $\hbar\omega$ the 
effective 
theory predicts a lower bound state off-set by about $0.02$ eV with 
respect to the full propagation matrix calculation (Fig. 
4(c)). This difference is of the order of the phonon energy and one order 
of magnitude lower than the bound state energy. However we notice that 
although the curves for the effective theory and numerical results are 
off, they present the same qualitative behavior. Therefore, 
one can affirm that the effective theory reproduces the numerical 
results with 
striking accuracy for small values of $g$. The same procedure can be 
repeated for any number of phonon channels with similar results.

Finally, let us turn to the current-voltage characteristics caused
by phonon assisted transmission features. As shown in Fig. 5(a), application
of an external electric field yields a spatial gradient in the 
potential energy profile.
The resulting current flow is determined from 
the transmission functions $T(E,V)$ at given voltage biases $V$
via an integral
\bea
I(V) = \int_0^{V_0} T(E,V) dE,
\eea
where the energy window $[0,V_0]$ for currents through semiconductor 
heterostructures is small compared to 
molecular junctions. As a result of this, the $I(V)$ dependence shown in 
Fig. 5(c) inherits the peak structure of the individual transmission 
curves, some of which are shown in Fig. 5(b). This is an important difference
from the step-like $I(V)$ curves reported in measurements of molecular
junctions.\cite{mitra04,braig03}

\begin{figure}[h]
\centering
\includegraphics[scale=0.57]{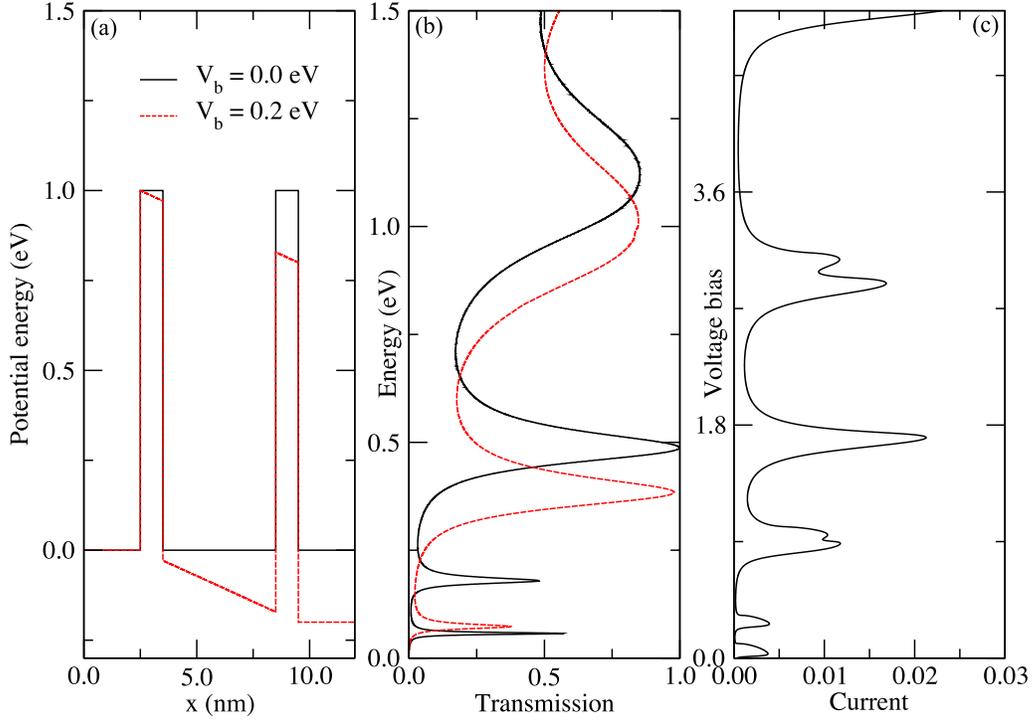}
\vspace{0cm}
\caption{\small 
(a) Potential double barrier with thickness of 1.0 nm and separation of
5.0 nm. The height of the barrier is 1.0 eV, and the electronic
effective mass is $0.07 m_e$. (b) Transmission vs. energy curves for one
excited phonon at the center of the potential well. In the
presence of a voltage bias across the heterostructure, this curve is
shifted towards lower energies. (c) Current-voltage
 curve for the double barrier in
(a). The two small peaks at low voltage bias ($V_b$) correspond to the
two low energy peaks in (b). The calculation of current is done by
integrating the transmission over an energy window from 0 eV to 50 meV.}
\end{figure}

\section{Conclusion}

In summary, we have investigated how quantum well electron-phonon 
resonances are affected by the presence of several inelastic channels in 
the phonon spectrum. Using a generalized propagation matrix method for 
multiple eleastic and inelastic transmission channels, we 
determined the highly non-perturbative effects of 
scattering by Einstein phonons on the electron transmission through 
potential structures. In particular, 
we observed a characteristic splitting of the 
bound state resonances into satellite peaks. The presence or absence
of a non-bonding resonance reflects the parity associated with even 
vs. odd numbers of accessible inelastic channels. Furthermore, in the 
limit of many available channels, the formation of low-energy 
impurity bands was observed. The dependence of the resonance satellites
on the electron-phonon coupling strength and on the phonon energies 
could be reproduced using an effective model, which works well 
within the limits of perturbation theory. 
One promise of the multi-channel propagation matrix method which is 
developed here lies in the ability to study highly asymmetric quantum 
systems with strongly interacting itinerant and local features. A 
further direction to pursue is to depart from strictly local oscillators,
which are nevertheless important for nanoelectronics, and to consider    
spatially extended phonon scattering regions.

\section{Acknowledgments}

We wish to thank A.F.J. Levi and B. Normand for useful comments, and acknowledge
support by the Department of Energy under grant DE-FG02-05ER46240.

\section{Appendix}

As an illustration of how to construct the 
generalized propagation matrix 
in practice, let 
us consider the example of a rectangular potential barrier given by  
\bea
V(x) = \left\{ 
\begin{array}{rl} 
0, &  x < a\\ 
V_0, &  a\le x \le b\\ 
0, &  x > b
\end{array} \right. 
\eea 
The wavefunctions are written 
as superpositions of plane waves, 
\bea
\psi_n(x<a) & = & a_ne^{ik_nx}+b_ne^{-ik_nx} , k_n=\sqrt{E-n\omega},  \\
\psi_n(x>b) & = & t_ne^{ik_nx} ,  k_n=\sqrt{E-n\omega}, \\
\psi_n(a\le x\le b) & = & c_ne^{-\kappa_nx}+d_ne^{\kappa_nx},  
\kappa_n=\sqrt{V_0-E-n\omega},
\eea
where $n$ 
represents the available phonon channels.  
It is assumed as an initial condition that the incident electrons enter the
potential structure from the left, i.e. $a_0 = 1, a_{n\ne 0} = 0$. 

To determine the transmission coefficients for the 
elastic ($t_{n=0}$) and inelastic ($t_{n\ne 0}$) channels
we use the propagation matrix 
method, matching the conditions $\psi_{n,j}=\psi_{n,j+1}$ and 
$d\psi_{n,j}/dx=d\psi_{n,j+1}/dx$ at each boundary. 
In the absence of phonon scattering ($n=0$),
the propagation matrix $\rho$ of the system is obtained by  
multiplication of the step matrices (at $x=a$ and $x=b$) and the free 
propagation matrix  
for $a\le x\le b$,
\begin{equation}
\hat\rho=\hat\rho_{step}^a\hat\rho_{free}^L\hat\rho_{step}^b,
\end{equation}
which are given by
\begin{equation}
\hat\rho_{step}^a=
\frac{1}{2} \left(
\begin{array}{cc} 
1+\frac{\kappa_{0}}{k_0} & 1-\frac{\kappa_{0}}{k_0}\\ 
1-\frac{\kappa_{0}}{k_0} & 1+\frac{\kappa_{0}}{k_0}
\end{array} \right)
\end{equation}
\begin{equation}
\hat\rho_{step}^b=
\frac{1}{2}\left(
\begin{array}{cc} 
1+\frac{k_0}{\kappa_0} & 1-\frac{k_0}{\kappa_0} \\ 
1-\frac{k_0}{\kappa_0} & 1+\frac{k_0}{\kappa_0}
\end{array} \right)
\end{equation}
And in the interval $a\le x\le b$:
\begin{equation}
\hat\rho_{free}^L=
\left(
\begin{array}{cc} 
e^{\kappa_0L} & 0 \\ 
0 & e^{-\kappa_0L} \\ 
\end{array} \right)
\end{equation}
with $L=b-a$. 

In order to find the transmission probability, we need to solve the matrix equation $\hat\rho\textbf{t}=\textbf{a}$ where the coefficients of the wave functions (7) and (8) are the elements of $\textbf{a}$ and  $\textbf{t}$, respectively:

\begin{equation}
\left(
\begin{array}{cc} 
\rho_{11} & \rho_{12}\\ 
\rho_{21} & \rho_{22}
\end{array} \right)\left(
\begin{array}{c}
t_0 \\ 0
\end{array} \right)=
\left(\begin{array}{c}
a_0 \\ b_0
\end{array} \right)
\end{equation}

For this system without phonon channels, the transmission probability $T(E)$ is simply:

\begin{equation}
|t_0|^2=\left|\frac{a_0}{\rho_{11}}\right|^2
\end{equation}

If instead phonon scatterer 
centers are present, we do not have the above condition on the derivative, but rather 
integrate Schr\"odinger's equation around $x=0$ (from $-\epsilon$ to 
$+\epsilon$). 
If for instance, we add a phonon scatterer at $x=a$ in the 
same potential barrier, we have for the step and free matrices are:

\begin{equation}
\hat\rho_{step}^a=
\frac{1}{2} \left(
\begin{array}{cccc} 
1+\frac{i\kappa_{0}}{k_0} & 1-\frac{i\kappa_{0}}{k_0} & \frac{igm}{k_0 \hbar^2} & 
\frac{igm}{k_0 \hbar^2} \\ 
1-\frac{i\kappa_{0}}{k_0} & 1+\frac{i\kappa_{0}}{k_0} & \frac{-igm}{k_0 \hbar^2} & 
\frac{-igm}{k_0 \hbar^2} \\
\frac{igm}{k_1 \hbar^2} & \frac{igm}{k_1 \hbar^2} & 1+\frac{i\kappa_{1}}{k_1} & 
1-\frac{i\kappa_{1}}{k_1}  \\ 
\frac{-igm}{k_1 \hbar^2} & \frac{-igm}{k_1 \hbar^2} & 1-\frac{i\kappa_{1}}{k_1} & 
1+\frac{i\kappa_{1}}{k_1} 
\end{array} \right)
\end{equation}

\begin{equation}
\hat\rho_{step}^b=
\frac{1}{2}\left(
\begin{array}{cccc} 
1+\frac{k_0}{\kappa_0} & 1-\frac{k_0}{\kappa_0} & 0 & 0\\ 
1-\frac{k_0}{\kappa_0} & 1+\frac{k_0}{\kappa_0} & 0 & 0\\
0 & 0 & 1+\frac{k_1}{\kappa_1} & 1-\frac{k_1}{\kappa_1} \\ 
0 & 0 & 1-\frac{k_1}{\kappa_1} & 1+\frac{k_1}{\kappa_1} 
\end{array} \right)
\end{equation}

\begin{equation}
\hat\rho_{free}^L=
\left(
\begin{array}{cccc} 
e^{\kappa_0L} & 0 & 0 & 0\\ 
0 & e^{-\kappa_0L} & 0 & 0 \\ 
0 & 0 & e^{\kappa_1L}  & 0\\ 
0 & 0 & 0 & e^{-\kappa_1L}
\end{array} \right)
\end{equation}

Once again we determine $T(E)$ by solving $\hat\rho\textbf{t}=\textbf{a}$. However, now we have two transmission coefficients $t_0$ and $t_1$, one for each phonon channel, and the transmission probability is given by:

\begin{equation}
T(E)=|t_0|^2+\frac{k_1(E)}{k_0(E)}|t_1|^2 .
\end{equation}

\end{document}